\documentclass[twoside]{dis04}
\def\runauthor{Akio Ogawa for the STAR Collaboration}
\def\shorttitle{Forward Physics at STAR}
\begin{document}

\title{
Forward Physics at STAR\\
Status of analysis on forward and mid rapidity correlation measurements
in p+p and d+Au
}

\author{Akio Ogawa for the STAR collaboration}

\address{Brookhaven National Laboratory\\
Upton NY 11973-5000, USA\\
E-mail: akio@bnl.gov }

\date{\today} 

\maketitle

\abstracts{Measurements of the production of high energy $\pi^0$ mesons at large
pseudorapidity, coincident with charged hadrons at mid-rapidity, for
proton+proton and deuteron+gold collisions at $\sqrt{s_{NN}}=200\ $GeV
are reported.  The p+p cross section for inclusive $\pi^0$ production
follows expectations from next-to-leading order perturbative QCD.  Both
the inclusive cross section and the di-hadron azimuthal correlations
are consistent with a model where parton showers supplement
leading-order pQCD.  A suppression of the back-to-back azimuthal
correlations was observed in d+Au, qualitatively consistent with
the gluon saturation picture.}


It has been established from deep inelastic scattering (DIS)
experiments that parton distribution functions (PDFs) are modified for
nucleons bound in heavy nuclei.  When the partons carry only a small
fraction of the bound nucleon's momentum (Bjorken $x<0.1$), nuclear
PDFs are found to be smaller than nucleon PDFs\cite{eks}.
This phenomenon is known as shadowing.  Establishing a
quantitative understanding of nuclear PDFs is an essential step towards
understanding the dynamics of relativistic heavy ion collisions to
address whether such collisions form a quark gluon plasma.  Knowledge
of nuclear gluon distribution functions is particularly important and
is, at present, limited because of the limited range in resolution
scale ($Q^2$) spanned by the world data on nuclear DIS.  Studies of
proton+nucleus collisions at large center of mass energies
($\sqrt{s_{NN}}$=200 GeV) can provide constraints to the gluon density
in heavy nuclei.

In the perturbative QCD explanation of large-rapidity particle
production, a large-$x$ parton, typically a quark, scatters from a
low-$x$ parton and then fragments into the
observed particle(s).  
Forward charged particle production is found to be suppressed in 
d+Au collisions \cite{brahms}, consistent with the expectation of
gluon saturation \cite{kharzeev-incl,jamal-2}, indicating a 
different mechanism for particle production.
Explanations of the suppression based on leading-twist perturbative
QCD calculations employing a model of gluon shadowing 
have also been suggested \cite{vogt}. Further tests of the
possible role played by gluon saturation at RHIC energies could be provided by
the study of particle correlations \cite{monojet}.  If the gluon
density in the incident heavy ion is saturated, the large-$x$ parton
from the deuteron is expected to undergo multiple
interactions through the dense gluon field resulting in multiple
recoil partons instead of a single recoil parton.  Hence, back-to-back di-hadron
azimuthal correlations for d+Au collisions are expected to be smaller
than those in p+p collisions.  
Coherent multiple parton scattering can also modify di-hadron azimuthal 
correlations \cite{qiuvitev} which may be another approach to particle 
production in the presence of a saturated gluon distribution.  
Leading twist calculations of particle
production including conventional shadowing do not result in
modifications of di-hadron azimuthal correlations \cite{hijing}.

An important question to address is whether fixed-order pQCD is
appropriate to describe forward particle production in p+p collisions
at $\sqrt{s}$=200 GeV.  For $\sqrt{s}\leq 62$ GeV, next-to-leading
order pQCD severely underpredicts measured $\pi^0$ cross sections
\cite{soffer}.  At $\sqrt{s}$=200 GeV and larger collision energies,
there is quantitative agreement between NLO pQCD calculations and
measured cross sections at mid-rapidity \cite{phenix}.  This agreement
has been found to extend to $\pi^0$ production at $\langle \eta
\rangle=3.8$ \cite{star-pi0}.  Further tests of the underlying
dynamics responsible for forward particle production can be obtained
from the study of particle correlations.  In particular, strong
azimuthal correlations of hadron pairs are expected when particle
production arises from $2 \rightarrow 2$ parton scattering, where in
the hard scattering two initial-state partons scatter and result in
two final-state partons.  
%
%
When a parton is scattered to fixed $\eta$, the pseudorapidity distribution
of the recoil parton is quite broad and is given by the convolution of PDFs
with the angular distribution for partonic scattering.
Extending pQCD beyond leading order introduces $2\rightarrow 3$ parton scattering 
%
that leads to a reduction of back-to-back azimuthal correlations of final-state hadron pairs.  
Similar effects are expected as the rapidity interval 
($\Delta\eta$) between jet pairs increases \cite{mueller}.

\begin{figure}
  \centerline {\includegraphics[width=10.0cm,clip]{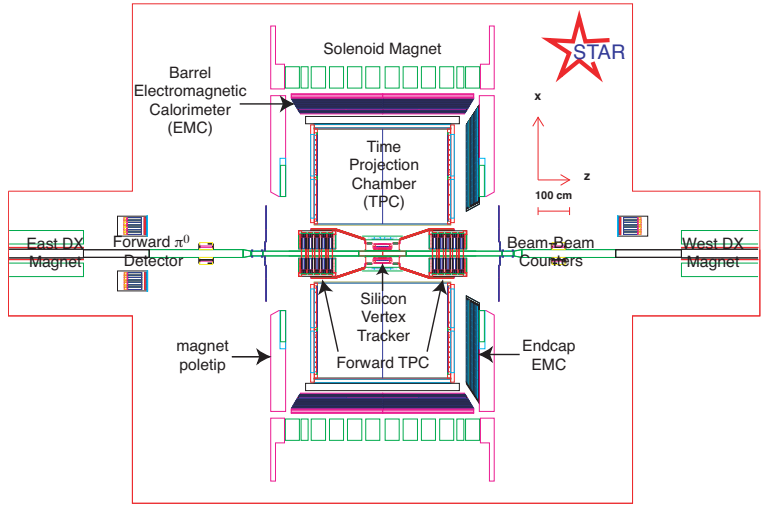}}
  \caption{\label{fig:star} Top view of the STAR detector.  The scale
  is such that displacements transverse to the beam ({\it x}) are two times
  larger than displacements along the beam ({\it z}).}

\end{figure}

This paper reports cross sections for forward inclusive $\pi^0$ production
for p+p collisions at $\sqrt{s}$=200 GeV. 
The azimuthal correlations between a forward $\pi^0$ ($\langle \eta_{\pi}
\rangle$=4.0) and mid-rapidity charged hadrons were studied.
These measurements for p+p collisions allow us to test our understanding 
of the particle production mechanism.
In addition, exploratory studies with d+Au collisions at $\sqrt{s_{NN}}$=200 GeV
are reported, and azimuthal correlations of hadron pairs separated by 
large $\Delta\eta$ were compared to those for p+p collisions.
Forward $\pi^0$ production in d+Au collisions refers to observation of the $\pi^0$ in the direction of the 
incident deuteron.


The Solenoidal Tracker At RHIC (STAR) is a multipurpose detector at
Brookhaven National Laboratory.  One of its principal components is a
time projection chamber (TPC) embedded in a 0.5T solenoidal magnetic
field used for tracking of charged particles produced at $|\eta|<$1.2.
A forward $\pi^0$ detector was installed in STAR as shown in Fig. \ref{fig:star} 
to allow for the detection of high energy $\pi^0$ mesons at large rapidity.
The FPD provides triggering and reconstruction of neutral pions produced 
with $3.3< \eta < 4.1$.  Data were collected over two years of RHIC operations.
In the 2002 run, p+p collisions were studied with a prototype FPD.
Details about triggering, event reconstruction and normalizations are
available in Ref. \cite{star-pi0}.  In the 2003 run, p+p collisions
were studied with the FPD and exploratory measurements were also
performed for d+Au collisions.

\begin{figure}
  \centerline{\includegraphics[width=5.7cm,clip]{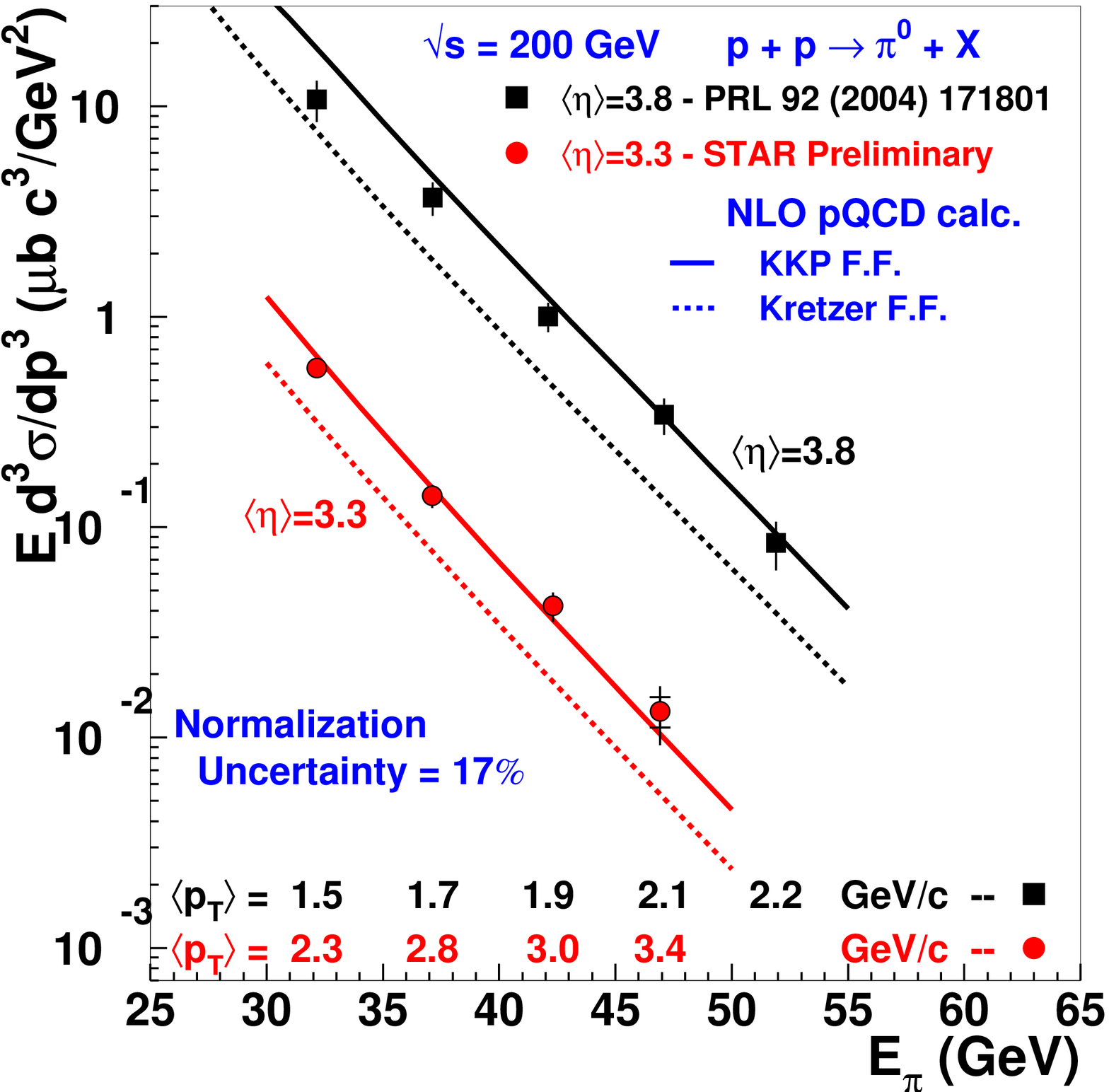}
  \includegraphics[width=6.2cm,clip]{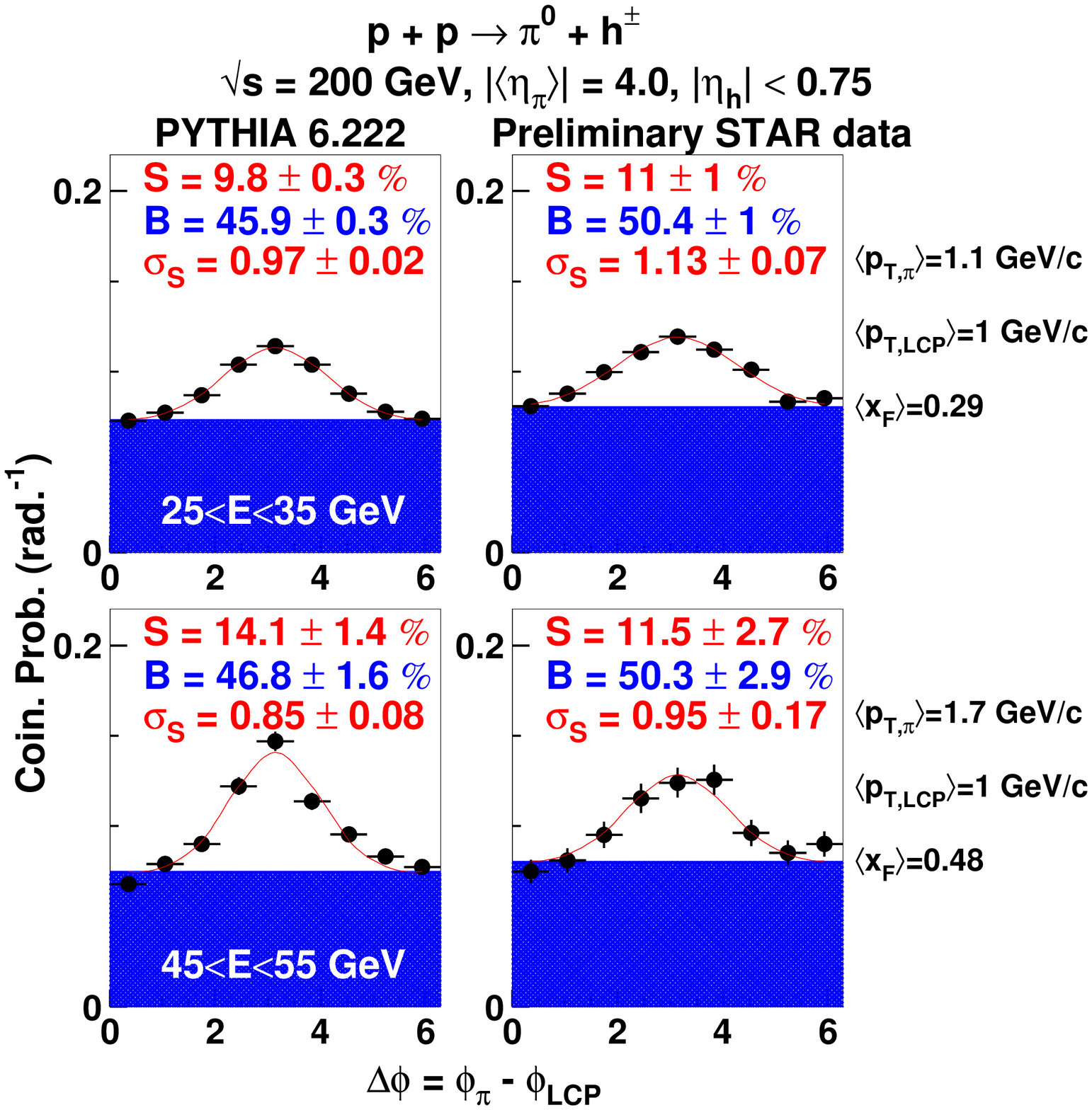}}
  \caption{\label{fig:crosssec} ({\it Left}) Inclusive $\pi^0$ production 
    cross section versus leading $\pi^0$ energy 
    ($E_\pi$) at average pseudorapidities ($\langle \eta \rangle$) 3.3 and 3.8.
    The average transverse momentum ($\langle p_T\rangle$) is correlated 
    with $E_\pi$ since the measurements were at a fixed $\eta$.
    The inner error bars are statistical, and are smaller 
    than the symbols for most data points.  The outer error bars
    combine these with $E_\pi$-dependent systematic errors added in 
    quadrature.  The curves are NLO pQCD calculations evaluated at 
    $\eta=3.3$ and 3.8 using different fragmentation functions as 
    described in the text.
  ({\it Right})
  Coincidence probability as function of azimuthal angle difference
  between $\pi^0$ at forward rapidity ($\langle \eta \rangle$=4.0) and the leading
  charged particle at mid rapidity ($|\eta|<0.75$, $p_T>0.5$ GeV),
  in two different $\pi^0$ energy bins for p+p collisions at
  $\sqrt{s}$=200 GeV. The left column is for PYTHIA simulations and the
  right column is data.  Fits to the data are described in the text.
  Error bars on the points represent only the statistical uncertainty.}

\end{figure}

The differential cross section for inclusive $\pi^0$ production for 
$30 < E_\pi < 55\ $GeV at $\langle \eta \rangle=3.8$ was previously reported \cite{star-pi0}.
The event reconstruction and normalization methods were extended to allow
measurement of the differential cross section at $\langle \eta \rangle=3.3$.  The
results are shown in Fig. \ref{fig:crosssec} in comparison to NLO
pQCD calculations evaluated at $\eta$=3.3 and 3.8 using CTEQ6M
\cite{cteq} parton distribution functions and equal renormalization
and factorization scales of $p_T$.  The NLO pQCD
calculations are consistent with the data, in contrast to
$\pi^0$ data at lower $\sqrt{s}$ \cite{soffer}.  The
solid line was calculated using the ``Kniehl-Kramer-P\"{o}tter'' (KKP) set of
fragmentation functions \cite{kniehl}, while the dashed line uses the
``Kretzer'' set \cite{kretzer}.  The difference between the two
reflects the uncertainty in the fragmentation functions at these
kinematics.  At the chosen scale, the KKP fragmentation functions tend
to agree with the data better than Kretzer, consistent with what has
been observed for midrapidity $\pi^0$ data at $\sqrt{s}=200\ $GeV
\cite{phenix}.

Correlations between a $\pi^0$ produced at large rapidity with large
Feynman $x$ ($x_F>0.25$) and charged particles produced at $|\eta|<0.75$ were also
studied.  
The efficiency uncorrected average multiplicity of charged particles with 
$|\eta|<$0.75 and $p_T >$0.2 GeV/c for events coincident with a forward 
$\pi^0$ is 5.1 for p+p, and 14.4 for d+Au.
This multiplicity is approximately 10\% larger than what is observed for 
minimum-bias events.  The leading
charged particle (LCP) analysis selects the mid-rapidity track with
the highest $p_T>$0.5 GeV/c in each event.  The azimuthal angle difference
$\Delta\phi=\phi_{\pi^0}-\phi_{LCP}$ is computed for each event.
Distributions of $\Delta\phi$, normalized by the number of $\pi^0$
observed at $\langle \eta_{\pi} \rangle$=4.0, for two
$\pi^0$ energy bins for p+p collisions are shown in
Fig. \ref{fig:crosssec}.  The left column of the figure shows simulations
using PYTHIA 6.222 \cite{pythia} that account for detector resolution
and reconstruction efficiency for both the forward $\pi^0$ and the mid-rapidity
charged particles.  The right hand column are preliminary STAR data.

The normalized $\Delta\phi$ distributions were fitted by the sum of a
constant and a Gaussian distribution centered at $\Delta\phi=\pi$.
Correlations near $\Delta\phi=0$ are not expected because of the large
rapidity interval between the $\pi^0$ and the LCP.  The fitted
parameters are the area under the Gaussian distribution (S),
representing the azimuthally correlated coincidence probability; the
azimuthally uncorrelated coincidence probability (B); and the width of
the Gaussian ($\sigma$), having contributions from transverse
momentum in the hadronization of the jets and
the momentum imbalance between the pair of jets ($k_T$).  The errors
on the fitted parameters are based on the full error matrix.

The PYTHIA simulation reproduces most features of the p+p
data.  The azimuthally correlated di-hadron coincidence probability (S) can be
identified in the PYTHIA simulation as arising from $2\rightarrow 2$ parton
scattering, resulting in a forward jet that fragments into the
observed forward $\pi^0$ and a midrapidity recoil jet that fragments
into the observed charged hadrons.  The width of the correlation may be
underestimated by PYTHIA indicating that the average $k_T$ between the
jet pair is too small.  The azimuthally uncorrelated coincidence probability
(B) primarily arises from $2\rightarrow 3$ partonic processes, 
fully accounted for in NLO pQCD calculations, and
approximately treated by initial- and final-state parton showers by
PYTHIA.  This interpretation is evident in the data when the azimuthal angle
used in the correlation analysis is derived from the vector sum of the
momenta of charged hadrons observed at midrapidity with $p_T>$0.2
GeV/c with a further condition that the magnitude of the vector sum
exceeds 0.5 GeV/c.  The azimuthally
correlated coincidence probability is found to be $\sim$ 30\% larger
in such an analysis with a corresponding reduction in B.

\begin{figure}
  \includegraphics[width=6.2cm,clip]{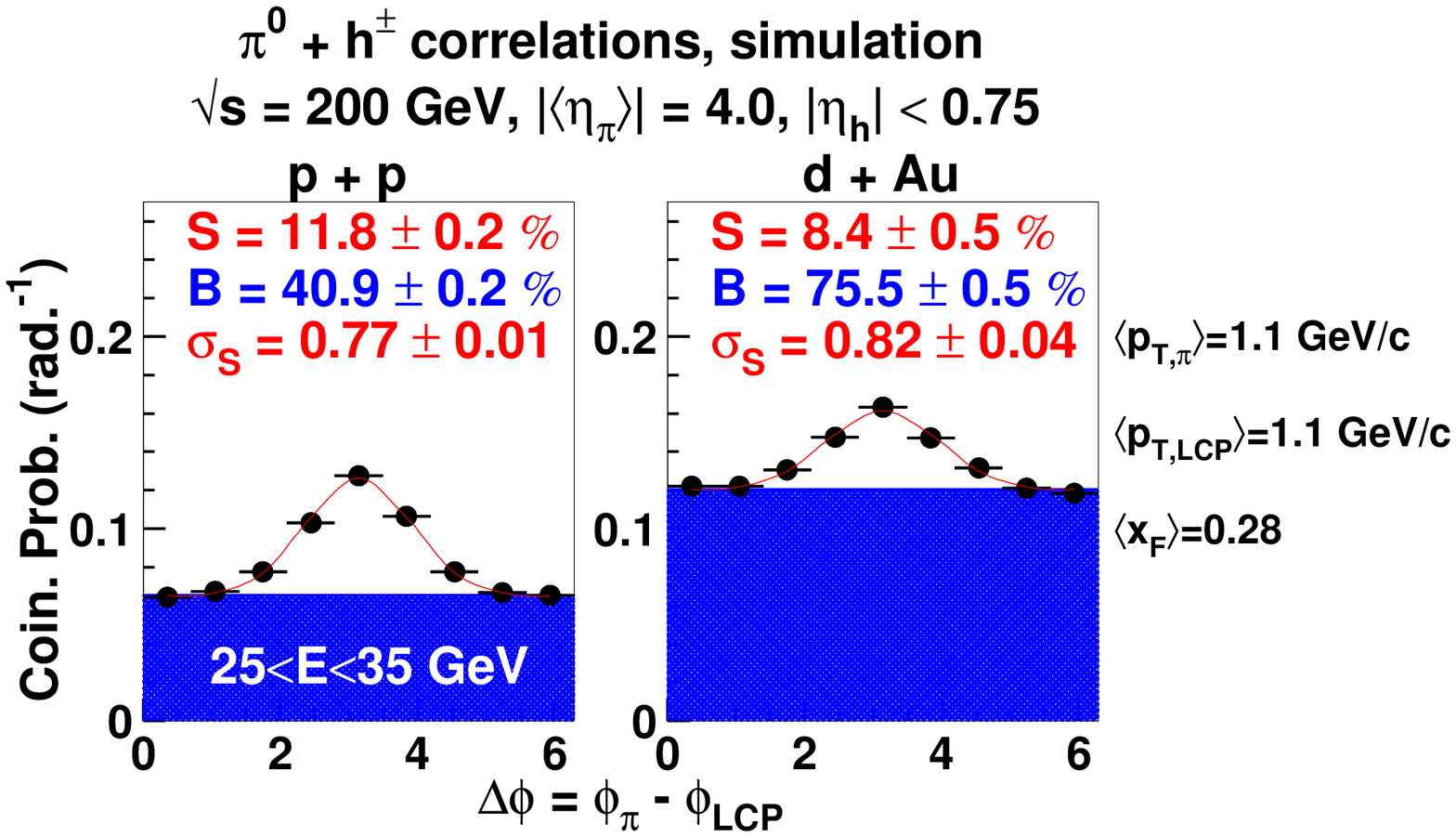}
  \includegraphics[width=6.2cm,clip]{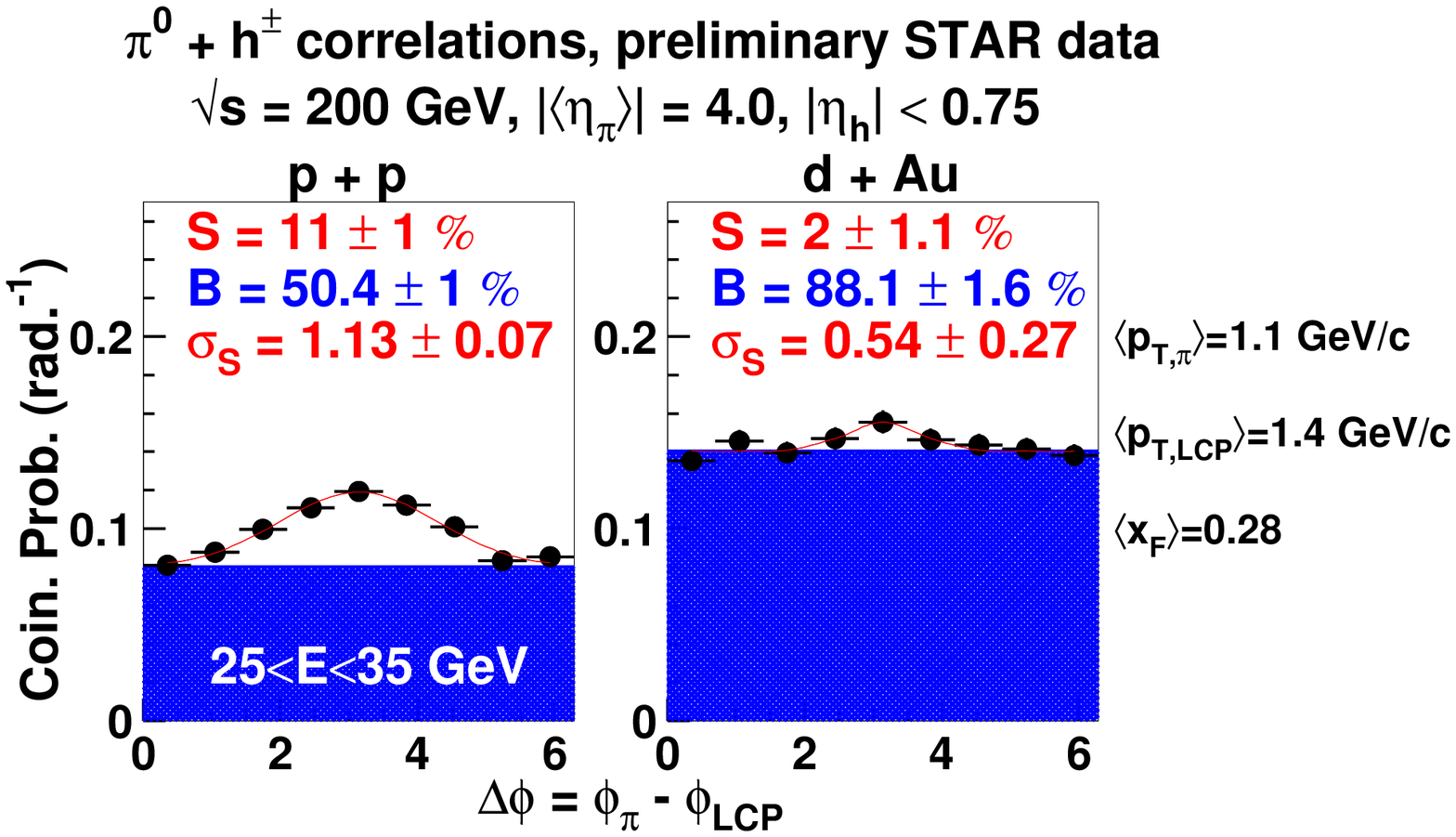}
  \caption{\label{fig:daucorr} 
  Coincidence probability as function of azimuthal angle 
  difference between $\pi^0$ at forward rapidity ($\langle \eta \rangle = 4.0$) and 
  leading charged particle at mid rapidity ($|\eta|<0.75$, $p_T>0.5$GeV).  
  ({\it Left}) HIJING 1.381 simulations for p+p and d+Au collisions, 
  including detector effects. HIJING has substantially smaller $k_T$
  compared to PYTHIA 6.222, and it gives smaller $\sigma$ in p+p compared 
  to PYTHIA (Fig. \ref{fig:crosssec}).
  ({\it Right}) STAR preliminary data for p+p and d+Au collisions.}
\end{figure}

An exploratory data set for forward $\pi^0$ production was obtained for
d+Au collisions.  For this data, the calibration of the FPD is, at
present, known to only 10\%.  This prevents an accurate determination
of the inclusive $\pi^0$ cross section for d+Au collisions.  Work is
underway to improve upon this calibration.  The slow variation of the
p+p$\rightarrow \pi^0+h^\pm$ correlations with $\pi^0$ energy suggests
that similar observables in d+Au collisions should be relatively
insensitive to calibration uncertainties.  The normalized $\Delta \phi$ 
distribution for d+Au collisions, analyzed in the same way as the p+p data,
is shown in comparison to p+p data in Fig. \ref{fig:daucorr}.  

There is a large increase of B$_{\rm dAu}$ relative to B$_{\rm
pp}$.  This increase is also expected in default HIJING simulations
\cite{hijing}, as shown in Fig. \ref{fig:daucorr}.
HIJING models d+Au collisions using PYTHIA for inelastic
nucleon-nucleon interactions and the Glauber model to account for
multiple collisions.
The growth in B arises when a nucleon from the deuteron beam interacts
with multiple nucleons from the Au beam.  The HIJING simulation does
not predict a significant difference between the width $\sigma_{\rm dAu}$
relative to $\sigma_{\rm pp}$.  For the data, the fitted $\sigma_{\rm dAu}$ is much
smaller than $\sigma_{\rm pp}$, most likely reflecting the inadequacy
of the functional form used to represent $\Delta\phi_{\rm dAu}$.  The
azimuthally correlated $\pi^0+h^\pm$ coincidence probability is smaller in d+Au
collisions than for p+p collisions.  Part of the reduction is due to
the fact that S is proportional to $\sigma$ and the fit results in a
width for $\Delta\phi_{\rm dAu}$ that is likely to be unphysically small. HIJING
simulations, that include a model of shadowing for nuclear PDFs, 
predict a significant azimuthally correlated $\pi^0+h^{\pm}$
coincidence probability.  This is not observed in the preliminary STAR data.

The largest systematic uncertainty in the $\Delta \phi$ distribution is expected to be the present
understanding of the calibration of the calorimeter.  Simulation
studies of the calorimeter response suggest that an improved
understanding of the calibration can be obtained.  The difference
between S$_{\rm pp}$ and S$_{\rm dAu}$ is observed in several
different methods of extracting an azimuthal angle from the charged
particles observed with $|\eta|<$0.75.  The impact of the functional
form used to represent the $\Delta\phi$ distribution is under
investigation.  Complete assessment of systematic
errors is underway.


In summary, cross sections for the inclusive production of $\pi^0$
mesons in p+p collisions at $\sqrt{s}$=200 GeV, at $\langle \eta_{\pi}
\rangle$=3.3 and 3.8 are found to be consistent with NLO pQCD
calculations, in contrast to the situation at smaller $\sqrt{s}$.
The azimuthal correlation between pairs of hadrons separated by large $\Delta\eta$ 
is described by a leading order pQCD calculation
including parton showers \cite{pythia} for p+p collisions.  Agreement of
these calculations with the inclusive cross sections and di-hadron correlations
strongly suggest that forward $\pi^0$ production arises from partonic
scattering at this collision energy.  This agreement means that forward particle production can
be exploited as a probe of low-$x$ parton densities in p+Au or d+Au
collisions at $\sqrt{s_{NN}}$=200 GeV.  Within the conventional
theoretical framework, the lowest $x$ values
can be reached when both hadrons are detected in the forward
direction \cite{gsv}.  The possible existence of gluon saturation in the Au
nucleus may modify the correlation between the recoil jet rapidity and
the initial-state Bjorken $x$ value.  Exploratory studies of forward
$\pi^0$ production in d+Au collisions suggest that the azimuthally
correlated component of hadron pairs separated by large rapidity
intervals is suppressed relative to what is observed for p+p
collisions and from model calculations.  This is qualitatively
consistent with the expectations that the particle production in a
dense gluon medium differs from conventional leading-twist NLO pQCD
expectations.  More data for forward particle production and di-hadron
correlations in d+Au collisions are required to reach a definitive
conclusion about the possible existence of a saturated gluon state in
the Au nucleus.  A quantitative theoretical understanding of the
rapidity and $p_T$ distribution of di-hadron correlations would
facilitate experimental tests of a possible color glass condensate.


\begin{thebibliography}{10}

  \bibitem{eks}
  K.~J.~Eskola, V.~J.~Kohinen and P.~V.~Ruuskanen,
  Nucl. Phys. {\bf B535},351 (1998);
  K.~J.~Eskola, V.~J.~Kolhinen and C.~A.~Salgado,
  Eur. Phys. J. C {\bf 9}, 61 (1999).

  \bibitem{brahms}
  I.~Arsene {\it et al.},
  {\mbox nucl-ex/0403005}.

  \bibitem{kharzeev-incl} Dmitri Kharzeev, Yuri V. Kovchegov and
  Kirill Tuchin, Phys.Rev. D {\bf 68} 094013 (2003).

  \bibitem{jamal-2} Jamal Jalilian-Marian, {\mbox nucl-th/0402080}.

  \bibitem{vogt} R. Vogt, {\mbox hep-ph/0405060}.

  \bibitem{monojet}
  D.~Kharzeev, E.~Levin and L.~McLerran,
  {\mbox hep-ph/0403271}.
             
  \bibitem{qiuvitev} Jianwei Qiu and Ivan Vitev, {\mbox hep-ph/0405068}.

  \bibitem{hijing}
  X.~N.~Wang and M.~Gyulassy, 
  Phys.\ Rev.\ D\ {\bf 44} 3501 (1991)

  \bibitem{soffer}  C.~Bourrely and J.~Soffer, 
  Eur. Phys. J. C {\bf 36}, 371 (2004) and {\mbox hep-ph/0311110}.

  \bibitem{phenix} S.\,S.~Adler {\it et al.}, Phys. Rev. Lett. {\bf
  91}, 241803 (2003).

  \bibitem{star-pi0}
  J.~Adams, {\it et al.},
  Phys. Rev. Lett. {\bf 92} 171801 (2004).

  \bibitem{mueller} A.H.~Mueller and H.~Navelet, 
  Nucl. Phys. {\bf B282} 727 (1987).

  \bibitem{cteq} J.~Pumplin {\it et al.}, J.\ High Energy Phys.\ 
    {\bf 0207}, 012 (2002).

  \bibitem{kniehl} B.\,A.~Kniehl {\it et al.}, Nucl.\ Phys.\ {\bf B597},
    337 (2001).

  \bibitem{kretzer} S.~Kretzer, Phys.\ Rev.\ D\ {\bf 62}, 054001 (2000).

  \bibitem{pythia} T.~Sj\"{o}strand, P.~Eden, C.~Friberg,
  L.~L\"{o}nnblad, G.~Miu, S.~Mrenna and E.~Norrbin,
  Comp. Phys. Commun. {\bf 135}, 238 (2001).

  \bibitem{gsv} V.~Guzey, M.~Strikman and W.~Vogelsang, {\mbox hep-ph/-0407201}.

\end{thebibliography}
\end{document}